\documentclass[11pt]{article}

\usepackage{amsmath,amssymb,graphicx}
\usepackage{hyperref}
\usepackage{tikz}
\usepackage{bm}
\usepackage{physics}
\usepackage{pgfplots}
\pgfplotsset{compat=1.17}

\title{Topological Gluon Mass and Shear Viscosity of the Quark--Gluon Plasma}

\author{
Debmalya Mukhopadhyay\\
\small Independent Researcher\\
\small Howrah, India\\
\small \texttt{debphys.qft@gmail.com}
}

\date{}

\newcommand{\pacs}[1]{\noindent\textbf{PACS numbers:} #1}

\begin{document}

\maketitle

\begin{abstract}

The quark--gluon plasma produced in relativistic heavy--ion
collisions behaves as a nearly perfect fluid characterized
by an exceptionally small shear viscosity to entropy density
ratio. Understanding the microscopic origin of this small
viscosity remains an important problem in the theory of
strongly interacting matter. In this work we investigate the transport properties of a
gluonic plasma in a non--Abelian gauge theory in which
gluons acquire a gauge--invariant mass through a
topological $B\wedge F$ interaction.
Integrating out the antisymmetric tensor field generates
an effective massive gluon propagator that modifies the
infrared behaviour of gluon exchange processes. Using relativistic kinetic theory and the Boltzmann
transport equation we compute the shear viscosity of the
plasma and derive the corresponding transport cross
section for gluon scattering.
The presence of the topological gluon mass provides a
natural infrared regulator for $t$--channel gluon exchange,
removing the divergence that appears in perturbative QCD
with massless gluons. We show that when the topological mass scale is comparable
to the soft momentum scale of the plasma, $m\sim gT$, the
resulting viscosity to entropy density ratio naturally
falls in the range inferred from hydrodynamic analyses of
heavy--ion collision experiments.
These results suggest that topological mass generation may
provide a simple microscopic mechanism contributing to the
near--perfect fluidity of the quark--gluon plasma.

\end{abstract}


\noindent

\section{Introduction}

Quantum chromodynamics (QCD) predicts that strongly interacting matter
under extreme conditions of temperature and density undergoes a phase
transition from a hadronic phase to a deconfined state of quarks and
gluons known as the quark--gluon plasma (QGP)
\cite{Shuryak2004,Gyulassy2005,Heinz2013}.
Such a state of matter is believed to have existed in the early universe
and can be recreated experimentally in relativistic heavy--ion
collisions at facilities such as the Relativistic Heavy Ion Collider
(RHIC) and the Large Hadron Collider (LHC).

One of the most remarkable properties of the QGP observed in heavy--ion
collisions is its strong collective hydrodynamic behaviour.
Experimental measurements of elliptic flow and other anisotropic flow
coefficients indicate that the plasma behaves as an almost perfect
fluid with extremely small viscosity
\cite{Romatschke2007,Song2011}.
Hydrodynamic simulations of heavy--ion collisions suggest that the
ratio of shear viscosity to entropy density lies in the range

\begin{equation}
0.08 \leq \frac{\eta}{s} \leq 0.20 ,
\end{equation}
which is remarkably close to the lower bound
\begin{equation}
\frac{\eta}{s} = \frac{1}{4\pi},
\end{equation}
derived for a large class of strongly coupled gauge theories with
gravity duals \cite{Kovtun2005,Son2007}.
The proximity of the QGP viscosity to this bound has generated
considerable interest in understanding the microscopic origin of this
strongly coupled behaviour.

In the weak coupling regime the shear viscosity of a gluonic plasma
can be calculated using relativistic kinetic theory and the Boltzmann
transport equation
\cite{Arnold2000,Arnold2003,Jeon1995,Jeon1996}.
Such perturbative calculations predict that the viscosity increases
rapidly as the gauge coupling becomes small, leading to values of
$\eta/s$ that are typically much larger than those inferred from
experimental data.
Understanding the origin of the small viscosity observed in the QGP
therefore remains an important theoretical problem.

A major difficulty in perturbative calculations of transport
coefficients in QCD arises from the infrared behaviour of gluon
exchange processes.
The dominant contribution to gluon scattering at high temperature
comes from $t$--channel exchange of nearly massless gluons.
This leads to infrared divergences in the scattering cross section,
which must be regulated by medium effects such as Debye screening
\cite{Braaten1990,Rebhan1993}.
In thermal Yang--Mills theory the infrared behaviour of gluon exchange
is typically treated using Hard Thermal Loop (HTL) resummation
\cite{Pisarski1984,Braaten1990}, which incorporates medium--induced
screening effects into the gluon propagator.

In the framework considered in the present work such resummation
is not required.
Instead, the infrared behaviour of the gauge theory is regulated
by a topological mechanism that generates a gauge--invariant mass
for the gluon field.
Specifically, the theory contains a topological $B\wedge F$
interaction that couples the Yang--Mills field strength to an
antisymmetric tensor field.
Integrating out the tensor field generates an effective mass term
for the gluon field while preserving gauge invariance.
The resulting gluon propagator is modified according to

\begin{equation}
\frac{1}{k^2} \rightarrow \frac{1}{k^2 - m^2},
\end{equation}
so that the mass parameter $m$ provides a natural infrared cutoff
for gluon exchange processes.

Topologically massive gauge theories in four dimensions have been
studied extensively in the context of gauge invariant mass generation
\cite{Freedman1981,Allen1991,Lahiri1997,Henneaux1997}.
In these theories the gauge boson mass arises from a topological
coupling between a one--form gauge field and a two--form tensor field,
rather than from spontaneous symmetry breaking.
This mechanism provides a natural infrared scale that can modify
the dynamics of the gauge theory. Such topological mass generation mechanisms are also
closely related to the appearance of a mass gap in the
infrared sector of non--Abelian gauge theories.
The existence of an infrared mass scale for gluonic
excitations has long been expected to play an important
role in the nonperturbative dynamics of QCD, including
confinement and the structure of the strongly interacting
vacuum.

The presence of a gluon mass changes the behaviour of scattering
amplitudes at small momentum transfer and therefore alters the
transport properties of the plasma. \medskip

In relativistic heavy--ion collisions the collective flow
observables measured at RHIC and the LHC are particularly
sensitive to the shear viscosity of the medium.
Since the shear viscosity is controlled by the microscopic
scattering rate of quasiparticles in the plasma, any mechanism
that modifies the infrared behaviour of gluon exchange
processes can have a direct impact on the hydrodynamic
properties of the quark--gluon plasma.
In particular, the topological mass regulates the infrared behaviour
of the $t$--channel gluon propagator and leads to a finite transport
cross section for gluon scattering. Transport coefficients of the quark--gluon plasma,
including shear viscosity and other dissipative
properties, have been extensively studied using
both perturbative and nonperturbative approaches
\cite{Moore2008,Schafer2009,Rischke2004}.

The purpose of the present work is to investigate whether
topological mass generation for gluons can provide a
microscopic mechanism for the small viscosity observed
in the quark--gluon plasma.
Using kinetic theory we compute the shear viscosity of the plasma
and determine how the modified gluon propagator affects the
scattering rate of gluons in the medium.
We further evaluate the ratio $\eta/s$ and examine whether the
presence of a topological gluon mass can naturally lead to values
compatible with those inferred from heavy--ion collision experiments.

The paper is organized as follows.
In Sec.~2 we review the topologically massive gauge theory and derive
the propagator of the massive gluon field.
In Sec.~3 we compute the gluon scattering amplitude and determine
the transport cross section in the presence of the topological mass.
In Sec.~4 we evaluate the shear viscosity using relativistic kinetic
theory.
In Sec.~5 we compute the ratio $\eta/s$ and discuss the implications
for the phenomenology of the quark--gluon plasma.
Finally, Sec.~6 contains a summary and outlook.

\section{Effective massive gluon propagator}

In this section we derive the propagator of the gluon field that arises
after integrating out the antisymmetric tensor field $B_{\mu\nu}$.
This procedure generates an effective gauge--invariant mass for the
gluon field.

\subsection{Quadratic part of the action}

To determine the propagators it is sufficient to consider the quadratic
part of the Lagrangian density\cite{Mukhopadhyay2020}

\begin{equation}
\mathcal{L} =
-\frac{1}{4}F_{\mu\nu}F^{\mu\nu}
+\frac{1}{12}H_{\mu\nu\lambda}H^{\mu\nu\lambda}
+\frac{m}{4}\epsilon^{\mu\nu\rho\lambda}B_{\mu\nu}F_{\rho\lambda}.
\end{equation}
For simplicity the non--Abelian color indices are suppressed in the
Lagrangian density. The gauge fields should be understood as
$A_\mu = A_\mu^a T^a$ and $B_{\mu\nu}=B_{\mu\nu}^a T^a$, where
$T^a$ are the generators of the SU($N_c$) gauge group.
Repeated color indices are implicitly summed.
The Yang--Mills field strength is
\begin{equation}
F_{\mu\nu}^a =
\partial_\mu A_\nu^a
-
\partial_\nu A_\mu^a
+
g f^{abc} A_\mu^b A_\nu^c .
\end{equation}
The field strength of the antisymmetric tensor field is

\begin{equation}
H^a_{\mu\nu\lambda} = (D_{[\mu} \,B_{\nu\lambda]})^a + g f^{abc} \,F_{[\mu\nu}^b\, C_{\lambda]}^c 
\end{equation}
where the fields $A^a_\mu$, $B^a_{\mu\nu}$ and $C^a_\mu$ are in the adjoint representation of the $SU(N_c)$ gauge group. To invert the kinetic operators we introduce gauge fixing terms

\begin{equation}
\mathcal{L}_{gf}
=
-\frac{1}{2\xi}(\partial_\mu A^\mu)^2
+
\frac{1}{2\eta}(\partial_\mu B^{\mu\nu})^2 .
\end{equation}
The quadratic action can then be written in momentum space.

\subsection{Momentum space representation}

After Fourier transformation

\begin{equation}
A_\mu(x) =
\int
\frac{d^4k}{(2\pi)^4}
e^{-ikx}A_\mu(k),
\end{equation}
and similarly for $B_{\mu\nu}$.
The quadratic Lagrangian becomes

\begin{equation}
\mathcal{L}
=
\frac{1}{2}
A_\mu(-k)
K^{\mu\nu}(k)
A_\nu(k)
+
\frac{1}{4}
B_{\mu\nu}(-k)
M^{\mu\nu,\rho\sigma}(k)
B_{\rho\sigma}(k)
+
\frac{im}{2}
\epsilon^{\mu\nu\rho\lambda}
B_{\mu\nu}(-k)
k_\rho
A_\lambda(k),
\end{equation}
where
\begin{equation}
K^{\mu\nu}(k)
=
k^2\eta^{\mu\nu}
-
k^\mu k^\nu .
\end{equation}
The tensor kinetic operator is

\begin{equation}
M^{\mu\nu,\rho\sigma}(k)
=
k^2
(\eta^{\mu\rho}\eta^{\nu\sigma}
-
\eta^{\mu\sigma}\eta^{\nu\rho})
+
\eta^{\mu\rho}k^\nu k^\sigma
-
\eta^{\mu\sigma}k^\nu k^\rho
-
\eta^{\nu\rho}k^\mu k^\sigma
+
\eta^{\nu\sigma}k^\mu k^\rho .
\end{equation}
\subsection{Integrating out the tensor field}
The partition function is
\begin{equation}
Z =
\int
\mathcal{D}A
\mathcal{D}B
\,
e^{iS[A,B]} .
\end{equation}
Since the action is quadratic in $B_{\mu\nu}$ we can integrate it out
exactly. The equation of motion for $B_{\mu\nu}$ is

\begin{equation}
M^{\mu\nu,\rho\sigma}
B_{\rho\sigma}
+
im
\epsilon^{\mu\nu\rho\lambda}
k_\rho
A_\lambda
=0.
\end{equation}
Solving for $B_{\mu\nu}$ gives

\begin{equation}
B_{\mu\nu}
=
-\frac{im}{k^2}
\epsilon_{\mu\nu\rho\lambda}
k^\rho
A^\lambda .
\end{equation}
Substituting this solution back into the action produces an effective
action depending only on $A_\mu$.

\subsection{Effective action}

The resulting effective quadratic Lagrangian becomes

\begin{equation}
\mathcal{L}_{eff}
=
-\frac14 F_{\mu\nu}F^{\mu\nu}
+
m^2
A_\mu
\left(
\eta^{\mu\nu}
-
\frac{k^\mu k^\nu}{k^2}
\right)
A_\nu .
\end{equation}
This term represents a gauge invariant mass term for the gluon. The mass term obtained after integrating out the tensor field is
transverse and therefore preserves gauge invariance.
Unlike the Proca mass term $m^2 A_\mu A^\mu$, the operator

\[
A_\mu
\left(
\eta^{\mu\nu}-\frac{\partial^\mu\partial^\nu}{\Box}
\right)
A_\nu
\]
projects onto the transverse component of the gauge field.
As a result the gauge symmetry of the original theory remains intact.

\subsection{Massive gluon propagator}

The inverse propagator of the gluon field therefore takes the form

\begin{equation}
D^{-1}_{\mu\nu}(k)
=
(k^2-m^2)\eta_{\mu\nu}
-
k_\mu k_\nu .
\end{equation}
Inverting this tensor yields the propagator

\begin{equation}
D_{\mu\nu}(k)
=
\frac{-i}{k^2-m^2}
\left(
\eta_{\mu\nu}
-
\frac{k_\mu k_\nu}{k^2}
\right)
-
i\xi
\frac{k_\mu k_\nu}{k^4}.
\end{equation}
This propagator describes a massive vector boson with mass $m$ while
preserving gauge invariance. The topological $B\wedge F$ interaction generates
additional interaction vertices involving the tensor
field $B_{\mu\nu}$ and the gauge field.
However, since the action is quadratic in $B_{\mu\nu}$,
the tensor field can be integrated out exactly.
The resulting effective theory contains a massive
gluon propagator while the explicit $B_{\mu\nu}$
degrees of freedom no longer appear.
Consequently the tree-level gluon scattering diagrams
are the same as in Yang–Mills theory, but with the
propagator modified to
\begin{equation}
\frac{1}{k^2-m^2}.
\end{equation}
The presence of this mass modifies the infrared behaviour of gluon
exchange processes and will play a crucial role in determining the
transport properties of the plasma. The propagator describes a massive vector excitation with
three physical polarization states. In covariant gauges the
propagator contains a gauge-dependent longitudinal component,
while the physical propagating modes correspond to the
transverse polarizations.
A massive vector boson possesses three physical
polarization states corresponding to two transverse
modes and one longitudinal mode.
However, in the high-temperature regime relevant for
the quark–gluon plasma the dominant contribution to
the transport cross section arises from the transverse
gluon polarizations.
The longitudinal contribution is suppressed by powers
of $m/E$ and will therefore be neglected in the
present analysis. This approximation is justified in the high--temperature
regime $T \gg m$ where the dominant contribution to the
transport cross section arises from transverse gluon
polarizations. Topologically massive gauge theories in four dimensions have been
studied extensively in the context of gauge invariant mass generation
\cite{Freedman1981,Allen1991,Lahiri1997,Henneaux1997,Mukhopadhyay2020}.

\section{Gluon scattering and transport cross section}

The scattering amplitudes are computed using the
standard Feynman rules of Yang--Mills theory
\cite{Peskin,WeinbergQFT}. In this section we calculate the gluon scattering amplitude in the
presence of the topological gluon mass derived in the previous section.
The resulting scattering cross section will determine the relaxation
time and ultimately the shear viscosity of the plasma.

\subsection{Kinematics of gluon scattering}

The dominant scattering process in a gluonic plasma is elastic
two–body scattering
\begin{equation}
g(p_1)+g(p_2)\rightarrow g(p_3)+g(p_4).
\end{equation}
The Mandelstam variables are defined as

\begin{equation}
s=(p_1+p_2)^2 ,
\end{equation}

\begin{equation}
t=(p_1-p_3)^2 ,
\end{equation}

\begin{equation}
u=(p_1-p_4)^2 .
\end{equation}
These variables satisfy
\begin{equation}
s+t+u = 4m^2
\end{equation}
Since the gluons acquire a mass $m$ through the
topological mechanism discussed in Sec.~II,
the Mandelstam variables satisfy
$s+t+u = 4m^2$ rather than the massless relation
$s+t+u=0$.
\subsection{Dominant scattering channel}

The leading contribution to gluon scattering at high temperature
arises from $t$–channel gluon exchange.
Using the massive propagator derived previously
\begin{equation}
D_{\mu\nu}(k)=
\frac{-i}{k^2-m^2}
\left(
g_{\mu\nu}-\frac{k_\mu k_\nu}{k^2}
\right),
\end{equation}
the propagator in the $t$ channel becomes

\begin{equation}
\frac{1}{t-m^2}.
\end{equation}
The corresponding matrix element can be written as
\begin{equation}
\mathcal{M}_t
=
g^2 f^{abe} f^{cde}
\,
\frac{1}{t-m^2}
\,
\epsilon_1^\mu
\epsilon_2^\nu
\epsilon_3^\rho
\epsilon_4^\sigma
\, N_{\mu\nu\rho\sigma},
\end{equation}
where $N_{\mu\nu\rho\sigma}$ represents the tensor structure arising from the triple–gluon vertices, 
\begin{equation}
N_{\mu\nu\rho\sigma}
=
\Gamma_{\mu\nu\alpha}(p_1,p_2,-k)
\left(
g^{\alpha\beta}-\frac{k^\alpha k^\beta}{k^2}
\right)
\Gamma_{\rho\sigma\beta}(-p_3,-p_4,k),
\end{equation}
with the triple–gluon vertex tensor
\begin{equation}
\Gamma_{\mu\nu\alpha}(p,q,r)
=
g_{\mu\nu}(p-q)_\alpha
+
g_{\nu\alpha}(q-r)_\mu
+
g_{\alpha\mu}(r-p)_\nu .
\end{equation}
The full tree-level gluon scattering amplitude
contains contributions from $s$-, $t$-, and $u$-channel
exchange diagrams together with the four-gluon
interaction vertex.
In a high-temperature plasma the dominant
contribution to momentum transport arises from
small-angle scattering where the momentum transfer
$|t|$ is much smaller than $s$.
In this kinematic regime the $t$-channel propagator
is strongly enhanced, while the $s$- and $u$-channel
contributions remain suppressed.
For this reason we retain only the $t$-channel
contribution in the calculation of the transport
cross section. \medskip

The dominance of the $t$--channel contribution can be
understood parametrically.
In the small--angle scattering regime relevant for
transport processes the momentum transfer satisfies
$|t| \ll s$.
The propagators appearing in the different channels
scale as
\begin{equation}
\frac{1}{t-m^2}, \qquad
\frac{1}{s-m^2}, \qquad
\frac{1}{u-m^2}.
\end{equation}
Since the typical center--of--mass energy in the plasma
is $s \sim T^2$ while the momentum transfer is of order
$|t| \sim g^2 T^2$, the ratio of propagators behaves as
\begin{equation}
\frac{s}{|t|} \sim \frac{1}{g^2} .
\end{equation}
For weak or moderately strong coupling this enhancement
makes the $t$--channel exchange the dominant contribution
to the scattering amplitude.

After averaging over initial colors and polarizations the squared
matrix element becomes
\begin{equation}
|\mathcal{M}|^2 =
\frac{9}{2} g^4
\frac{s^2+u^2}{(t-m^2)^2}.
\end{equation}
\medskip
Because the gluons acquire a topological mass, the vector field
in principle possesses three physical polarization states.

The completeness relation for a massive vector boson is

\begin{equation}
\sum_\lambda
\epsilon_\mu^{(\lambda)}(p)
\epsilon_\nu^{(\lambda)*}(p)
=
-
\left(
g_{\mu\nu}
-
\frac{p_\mu p_\nu}{m^2}
\right).
\end{equation}
However, in the high--temperature regime relevant for the
quark--gluon plasma the typical gluon energy satisfies
$E \sim T$, while the mass scale is assumed to be
$m \sim gT$.
The contribution of the longitudinal polarization
to the scattering amplitude is therefore suppressed by
powers of $m^2/E^2 \sim g^2$.
To leading order in the weak--coupling expansion the
dominant contribution to the transport cross section
arises from transverse gluon polarizations, and the
longitudinal contribution can be neglected.

\subsection{Differential cross section}

The differential cross section for two–body scattering is

\begin{equation}
\frac{d\sigma}{dt}
=
\frac{|\mathcal{M}|^2}{16\pi s^2}.
\end{equation}
Substituting the matrix element gives

\begin{equation}
\frac{d\sigma}{dt}
=
\frac{9 g^4}{32\pi}
\frac{s^2+u^2}{s^2(t-m^2)^2}.
\end{equation}
Using the Mandelstam relation for massive gluons

\begin{equation}
s+t+u = 4m^2 ,
\end{equation}
the variable $u$ can be written as

\begin{equation}
u = 4m^2 - s - t .
\end{equation}
In the kinematic regime relevant for a high--temperature
plasma the typical center--of--mass energy satisfies
$s \gg m^2$ and $s \gg |t|$.
In this limit the mass corrections to the Mandelstam
variables are negligible and one may approximate

\begin{equation}
u \approx -s .
\end{equation}

Consequently

\begin{equation}
s^2+u^2 \approx 2s^2 .
\end{equation}
The detailed evaluation of the transport cross section is presented in Appendix A and B.
The result can be written as
\begin{equation}
\sigma_{\mathrm{tr}}
=
\frac{9 g^4}{32\pi p^2}
\ln\!\left(\frac{s+m^2}{m^2}\right).
\end{equation}
In a thermal plasma the typical center--of--mass
energy satisfies $s \sim T^2$.
The transport cross section therefore becomes
\begin{equation}
\sigma_{\mathrm{tr}}
\approx
\frac{9 g^4}{32\pi T^2}
\ln\!\left(\frac{T^2}{m^2}\right).
\end{equation}

\subsection{Physical interpretation}

In perturbative QCD with massless gluons the transport cross
section contains an infrared divergence arising from the
$t$–channel exchange of nearly massless gluons.
The differential cross section behaves as
\begin{equation}
\frac{d\sigma}{dt}\propto \frac{1}{t^2},
\end{equation}
which leads to a logarithmic divergence in the transport cross section at small momentum transfer.

In the present theory the gluon propagator is modified by the presence of the topological mass
\begin{equation}
\frac{1}{t} \rightarrow \frac{1}{t-m^2}.
\end{equation}
The mass parameter $m$ therefore acts as an infrared cutoff that regulates the divergence of the scattering amplitude.
As a result the transport cross section becomes finite and takes the form
\begin{equation}
\sigma_{\mathrm{tr}}
=
\frac{9 g^4}{32\pi T^2}
\ln\!\left(\frac{T^2}{m^2}\right).
\end{equation}

\medskip

It is useful to compare this behaviour with the well-known
leading-order kinetic theory result for the shear viscosity
of a weakly coupled gluon plasma derived by Arnold, Moore
and Yaffe \cite{Arnold2000,Arnold2003}. In perturbative QCD
the dominant contribution to the transport cross section
also arises from $t$--channel gluon exchange and leads to a
logarithmic enhancement associated with the infrared region
of momentum transfer. In that case the infrared divergence
is regulated by medium effects incorporated through the
Hard Thermal Loop effective theory, which introduces a
screening scale of order $m_D \sim gT$ in the gluon
propagator. In the present framework the same infrared
behaviour is regulated instead by the topological gluon
mass $m$, which therefore plays a role analogous to the
screening scale in thermal QCD.

\section{Shear viscosity of the gluonic plasma}
In this section we compute the shear viscosity of the gluonic plasma
using relativistic kinetic theory.
Transport coefficients in a thermal plasma can be calculated
using the Boltzmann equation within the framework of
finite--temperature field theory \cite{KapustaBook,LaineBook}.
The viscosity is determined by the rate of momentum
transport carried by quasiparticles in the medium.

\subsection{Kinetic theory expression}

For a relativistic bosonic gas the shear viscosity can be written as

\begin{equation}
\eta =
\frac{4}{15T}
\int
\frac{d^3p}{(2\pi)^3}
\frac{p^4}{E_p^2}
\tau(p)
f_0(p)
\left[1+f_0(p)\right],
\end{equation}
where
\begin{equation}
E_p = \sqrt{p^2+m^2},
\end{equation}
is the quasi–particle energy and
\begin{equation}
f_0(p)=\frac{1}{e^{E_p/T}-1},
\end{equation}
is the Bose–Einstein distribution function.
The shear viscosity can be computed from the Boltzmann
equation describing the evolution of the gluon distribution
function in the plasma.
In the present work we employ the relaxation time
approximation, in which the collision integral is written
as
\begin{equation}
C[f] \approx -\frac{f-f_0}{\tau}.
\end{equation}
This approximation captures the dominant contribution
to momentum transport and is commonly used in kinetic
theory estimates of transport coefficients.
\subsection{Relaxation time}

The relaxation time is related to the scattering rate of
gluons in the plasma. For binary collisions the inverse
relaxation time can be written as

\begin{equation}
\tau^{-1}= n\,\langle \sigma_{\mathrm{tr}} v_{\mathrm{rel}} \rangle ,
\end{equation}
where $v_{\mathrm{rel}}$ is the relative velocity of the
colliding particles. For ultrarelativistic gluons in a hot
plasma the relative velocity is approximately unity,
$v_{\mathrm{rel}}\approx 1$, so that the relaxation time
reduces to
\begin{equation}
\tau^{-1}\approx n\,\sigma_{\mathrm{tr}} .
\end{equation}
The gluon number density is

\begin{equation}
n =
g_g
\int
\frac{d^3p}{(2\pi)^3}
\frac{1}{e^{p/T}-1},
\end{equation}
where
\begin{equation}
g_g = 2(N_c^2-1)
\end{equation}
is the number of gluonic degrees of freedom. The momentum integral can be evaluated using the standard Bose integral

\begin{equation}
\int_0^\infty
\frac{x^2\,dx}{e^x-1}
=
2\zeta(3).
\end{equation}
Changing variables $p=Tx$ gives

\begin{equation}
n=
g_g
\frac{\zeta(3)}{\pi^2}
T^3 .
\end{equation}
Substituting the transport cross section derived earlier
\begin{equation}
\sigma_{\mathrm{tr}}\approx
\frac{9 g^4}{32\pi T^2}
\ln\!\left(\frac{T^2}{m^2}\right).
\end{equation}
the relaxation time becomes
\begin{equation}
\tau\approx
\frac{32\pi T^2}
{9 g^4 n \ln(T^2/m^2)} .
\end{equation}
Substituting the expression for the number density yields
\begin{equation}
\tau \approx
\frac{32\pi^3}
{9 g^4 g_g \zeta(3)}
\frac{1}{T
\ln\!\left(\frac{T^2}{m^2}\right)} .
\end{equation}

\subsection{Evaluation of the viscosity integral}

We now evaluate the momentum integral appearing in the viscosity formula. For temperatures much larger than the gluon mass the energy can be
approximated as

\begin{equation}
E_p \approx p.
\end{equation}
The viscosity expression becomes
\begin{equation}
\eta=
\frac{4\tau}{15T}
\int
\frac{d^3p}{(2\pi)^3}
p^2
f_0(p)
\left[1+f_0(p)\right].
\end{equation}
The integral
\begin{equation}
I=
\int
\frac{d^3p}{(2\pi)^3}
p^2
f_0(p)
\left[1+f_0(p)\right]
\end{equation}
can be evaluated using the identity

\begin{equation}
f_0(1+f_0)=
\frac{e^{p/T}}{(e^{p/T}-1)^2}.
\end{equation}
Changing variables to $x=p/T$ gives
\begin{equation}
I=
\frac{T^5}{2\pi^2}
\int_0^\infty
\frac{x^4 e^x}{(e^x-1)^2}dx .
\end{equation}

Using the standard integral
\begin{equation}
\int_0^\infty
\frac{x^4 e^x}{(e^x-1)^2}dx
=
\frac{4\pi^4}{15},
\end{equation}
we obtain
\begin{equation}
I=
\frac{2\pi^2}{15}T^5 .
\end{equation}
Substituting this result into the viscosity formula yields
\begin{equation}
\eta=
\frac{8\pi^2}{225}\tau T^4.
\end{equation}

\subsection{Final expression}

Finally inserting the relaxation time gives

\begin{equation}
\eta\approx
\frac{256\pi^5}{2025}
\frac{T^3}
{g^4 g_g \zeta(3)\ln(T^2/m^2)} .
\end{equation}
The logarithmic factor $\ln(T^2/m^2)$ originates from the
integration over the small momentum transfer region of the
$t$--channel gluon exchange. The presence of the topological
mass $m$ therefore regulates the infrared behaviour of the
scattering amplitude and renders the transport cross section
finite.  As a result the viscosity remains finite and is controlled by the infrared
scale set by the topological gluon mass.

\subsection{Estimate of the topological mass scale}

In the previous sections the gluon mass parameter $m$
was treated as a phenomenological parameter generated
by the topological $B\wedge F$ interaction.
It is useful to estimate the expected magnitude of this
mass scale in a thermal plasma.

In finite--temperature gauge theory the propagation
of gauge bosons is modified by medium effects encoded
in the thermal gluon self--energy
\cite{Mukhopadhyay2020,KapustaBook,LaineBook}.
At leading order the longitudinal component of the
polarization tensor generates a screening scale
known as the Debye mass.
For an SU($N_c$) gauge theory the Debye mass is given by

\begin{equation}
m_D^2 =
\frac{N_c}{3}\, g^2 T^2 ,
\end{equation}
which follows from the one--loop evaluation of the
gluon polarization tensor in the thermal background
\cite{KapustaBook,LaineBook}.

The origin of this scale can be understood from the
hard thermal loop contribution to the gluon self--energy.
Schematically one obtains

\begin{equation}
\Pi_{\mu\nu}(k)
\sim
g^2
\int
\frac{d^3p}{(2\pi)^3}
\frac{1}{p}
f_B(p),
\end{equation}
where $f_B(p)$ is the Bose--Einstein distribution.
Since the thermal integral is dominated by momenta
of order $p\sim T$, dimensional analysis gives

\begin{equation}
\Pi_{\mu\nu}(k)
\sim g^2 T^2 .
\end{equation}
This result implies that thermal gauge theories
naturally generate an infrared momentum scale
of order

\begin{equation}
m_D \sim gT .
\end{equation}
In the present framework the gluon mass arises from
the topological $B\wedge F$ interaction rather than
from thermal loop corrections.
Nevertheless, the infrared behaviour of gluon
exchange processes in the plasma is controlled by
the same soft momentum scale.
It is therefore natural to assume that the magnitude
of the topological gluon mass is comparable to this
scale,

\begin{equation}
m \sim gT .
\end{equation}
Substituting this estimate into the expression for the viscosity obtained in the previous section gives the parametric behaviour
\begin{equation}
\eta
\sim
\frac{T^3}{g^4 \ln(1/g)} .
\end{equation}
Consequently the ratio of shear viscosity to entropy
density behaves as
\begin{equation}
\frac{\eta}{s}
\sim
\frac{1}{g^2 \ln(1/g)} .
\end{equation}
For moderately strong coupling $g\sim2$ this leads
to values of $\eta/s$ of order $0.1$, which is
consistent with phenomenological estimates extracted
from relativistic heavy--ion collision experiments.

Although this value of the coupling corresponds to a
moderately strongly interacting plasma, perturbative
kinetic theory calculations of QCD transport
coefficients are known to remain qualitatively
reliable in this regime
\cite{Arnold2000,Arnold2003,Schafer2009}.

\section{Viscosity to entropy density ratio}

In this section we evaluate the ratio of the shear viscosity to the
entropy density of the gluonic plasma. This quantity is of particular
importance in the phenomenology of the quark–gluon plasma because it
controls the strength of viscous effects in relativistic hydrodynamic
evolution.

\subsection{Entropy density of the gluonic plasma}

The entropy density can be obtained from the thermodynamic relation

\begin{equation}
s = \frac{\partial P}{\partial T},
\end{equation}
where $P$ is the pressure of the system.

For a relativistic bosonic gas the pressure is

\begin{equation}
P =
g_g
\int
\frac{d^3p}{(2\pi)^3}
\frac{p}{3}
\frac{1}{e^{p/T}-1}.
\end{equation}

Using the standard Bose integral

\begin{equation}
\int_0^\infty
\frac{x^3 dx}{e^x-1}
=
\frac{\pi^4}{15},
\end{equation}

the pressure becomes

\begin{equation}
P =
\frac{g_g \pi^2}{90}T^4 .
\end{equation}
The entropy density is therefore
\begin{equation}
s =
\frac{dP}{dT}
=
\frac{4 g_g \pi^2}{90}T^3 .
\end{equation}

Thus
\begin{equation}
s =
\frac{2 g_g \pi^2}{45}T^3 .
\end{equation}

\subsection{Ratio $\eta/s$}

Using the expression for the shear viscosity derived in the previous
section

\begin{equation}
\eta =
\frac{256\pi^5}{2025}
\frac{T^3}
{g^4 g_g \zeta(3)\ln(T^2/m^2)} .
\end{equation}
the ratio of shear viscosity to entropy density becomes
\begin{equation}
\frac{\eta}{s}
=
\frac{128\pi^3}{45 g_g \zeta(3)}
\frac{1}{g^4\ln(T^2/m^2)} .
\end{equation}

\subsection{Physical interpretation}

The viscosity depends logarithmically on the topological
mass parameter through the factor $\ln(T^2/m^2)$ which
acts as an infrared regulator for gluon scattering.
In perturbative QCD the absence of a gluon mass leads to an infrared
divergence in the transport cross section and the viscosity becomes
very large. In contrast, the topological gluon mass provides a natural infrared
cutoff which enhances the scattering rate of gluons in the plasma.
As a result the viscosity is reduced. The gluon mass parameter appearing in the present
framework originates from the topological $B\wedge F$ interaction and is therefore present already at tree-level in the effective gauge theory.
This mass should not be identified directly with the
Debye screening mass generated by thermal loop
corrections in QCD.

Nevertheless, in a thermal plasma the dominant
infrared momentum scale governing gluon exchange
processes is of order $gT$.
It is therefore natural to assume that the magnitude
of the topological gluon mass is comparable to this
scale,
\begin{equation}
m \sim gT .
\end{equation}

Substituting this estimate into the expression for
$\eta/s$ obtained above shows that the viscosity to
entropy density ratio remains of order
\begin{equation}
\frac{\eta}{s} \sim \frac{1}{g^2},
\end{equation}
up to logarithmic corrections.
For moderately strong coupling $g \sim 2$ this leads
naturally to values of $\eta/s$ of order $0.1$,
consistent with phenomenological estimates of the
quark--gluon plasma.

\begin{figure}[h!]
\begin{center}
\includegraphics[scale=0.48]{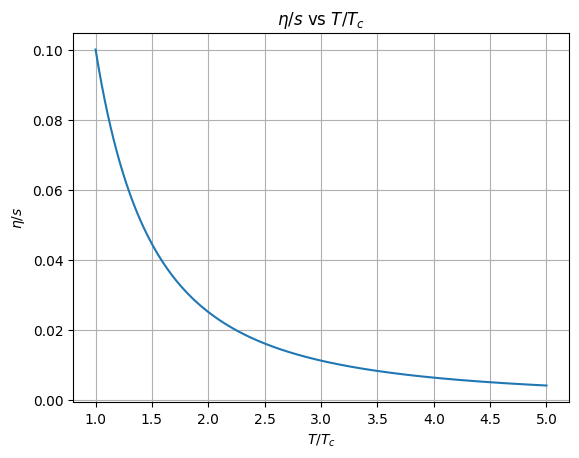}
\caption{Ratio of shear viscosity to entropy density as a function of
$m/T$ for coupling $g=2$ and $N_c=3$.}
\label{fig:etas_TTc}
\end{center}

\end{figure}

This result indicates that the presence of a topological gluon mass
can naturally produce a small viscosity to entropy ratio for moderate
values of the coupling constant.
\begin{equation}
\frac{\eta}{s}\approx 0.1 .
\end{equation}
The temperature dependence of the ratio $\eta/s$
can be illustrated by plotting the result as a
function of $T/T_c$, where $T_c$ denotes the
critical temperature of the deconfinement transition.
The resulting behaviour is shown in Fig.~\ref{fig:etas_TTc}.

\subsection{Implications for the quark–gluon plasma}
\begin{figure}[h!]
\begin{center}
\includegraphics[scale=0.48]{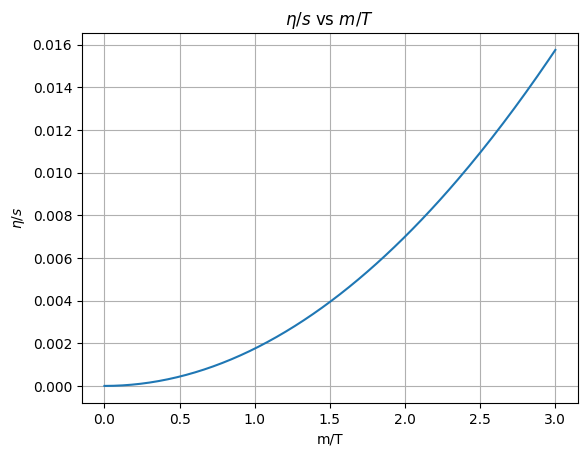}
\caption{Temperature dependence of $\eta/s$ as a function of $T/T_c$
for $g=2$.}
\label{fig:etas_mT}
\end{center}
\end{figure}

Hydrodynamic analyses of heavy–ion collision experiments indicate that
the quark–gluon plasma created in relativistic nuclear collisions
behaves as a nearly perfect fluid with a viscosity to entropy ratio
close to
\begin{equation}
\frac{\eta}{s}\approx 0.1 .
\end{equation}
To illustrate the dependence of the viscosity to entropy density ratio
on the topological gluon mass, it is useful to express the result in
terms of the dimensionless parameter $m/T$.
Using the expression derived above, the ratio $\eta/s$ can be written
as a function of $m/T$.
The resulting behaviour is shown in Fig.~\ref{fig:etas_mT}.

As shown in Fig.~\ref{fig:etas_mT}, the ratio $\eta/s$ increases
with the parameter $m/T$.
For values of the gluon mass comparable to the temperature,
the resulting viscosity remains small, indicating that the plasma
behaves as a nearly perfect fluid.
This behaviour is consistent with the phenomenological
observations of the quark--gluon plasma created in
relativistic heavy--ion collisions.
The plot indicates that the ratio $\eta/s$
takes its smallest value near the transition
temperature and increases as the temperature
moves away from $T_c$.
This behaviour is consistent with the picture
of the quark–gluon plasma as a nearly perfect
fluid near the deconfinement transition.

It is also interesting to compare the obtained values of the ratio
$\eta/s$ with the well-known Kovtun--Son--Starinets bound,
\begin{equation}
\frac{\eta}{s} \ge \frac{1}{4\pi},
\end{equation}
which was originally derived using the gauge/gravity duality.
Hydrodynamic analyses of heavy-ion collision experiments suggest
that the quark--gluon plasma produced in relativistic nuclear
collisions has a viscosity to entropy density ratio close to this
bound. The results obtained in the present framework indicate that
the presence of a topological gluon mass can naturally lead to
values of $\eta/s$ in this phenomenologically relevant range.

The results obtained in the present work indicate that
topological mass generation in non--Abelian gauge theories
can provide a natural infrared regulator for gluon exchange
processes in a hot plasma. When the topological mass scale
is comparable to the soft momentum scale of the medium,
$m \sim gT$, the resulting transport cross section leads
to a shear viscosity to entropy density ratio that lies
within the range extracted from hydrodynamic analyses of
relativistic heavy--ion collision experiments.

The present analysis therefore suggests that topological
mass generation may constitute a simple microscopic
mechanism contributing to the near--perfect fluid behaviour
of the quark--gluon plasma. A more detailed understanding of this mechanism will
require a systematic treatment of
finite--temperature effects and the inclusion of quark
degrees of freedom in the transport calculation. These
questions will be explored in future work.

\section{Conclusions}

In this work we have investigated the transport properties of a gluonic
plasma in a non–Abelian gauge theory where the gluons acquire a mass
through a topological $B\wedge F$ coupling. We first derived the effective massive gluon propagator obtained by
integrating out the antisymmetric tensor field. The resulting propagator
contains a gauge–invariant mass parameter which modifies the infrared
structure of gluon exchange processes.

Using this propagator we computed the scattering amplitude for
gluon–gluon interactions and derived the corresponding transport cross
section. The presence of the gluon mass removes the infrared divergence
which appears in perturbative QCD due to long–range gluon exchange. We then applied relativistic kinetic theory to calculate the shear
viscosity of the gluonic plasma. The resulting viscosity depends on the
ratio of the gluon mass scale to the temperature of the plasma and is
finite due to the infrared cutoff provided by the mass term.

The ratio of shear viscosity to entropy density was also evaluated.
We found that the quantity $\eta/s$ is controlled by the dimensionless
parameter $m^2/(g^4 T^2)$.
For values of the gluon mass comparable to the temperature, the ratio
$\eta/s$ can take values close to those extracted from hydrodynamic
analyses of relativistic heavy–ion collision experiments.

Our results therefore suggest that topological mass generation in
non–Abelian gauge theories may provide a microscopic mechanism for the
near–perfect fluid behaviour of the quark–gluon plasma. The present mechanism differs from the usual perturbative QCD
treatment in that the infrared behavior of gluon exchange
is regulated by a topological mass rather than by
Hard Thermal Loop resummation. An important feature of the present mechanism is that
the infrared behaviour of gluon exchange processes is
regulated already at the level of the effective gauge
theory through the topological mass term.
In contrast to the conventional perturbative treatment
of QCD, where infrared divergences are regulated by
thermal screening effects incorporated through Hard
Thermal Loop resummation, the present framework
provides an intrinsic infrared cutoff in the gluon
propagator.
This suggests that topological mass generation may
provide an alternative description of infrared
dynamics in the quark--gluon plasma.
This provides a conceptually simple microscopic origin
for the small viscosity of the quark--gluon plasma. \medskip

It would be interesting to explore whether the presence of a
topological gluon mass could be investigated using lattice
gauge theory. Lattice studies of the gluon propagator at
finite temperature provide information on the infrared
behaviour of gauge fields in the quark--gluon plasma.
In particular, an effective mass scale for gluonic
excitations can be extracted from the infrared structure
of the propagator. A comparison between such lattice
determinations of the finite--temperature gluon propagator
and the massive propagator predicted by the present
framework could provide a useful test of the role of
topological mass generation in the dynamics of the
quark--gluon plasma.

Several extensions of the present work are possible. In particular it
would be interesting to investigate the temperature dependence of the
topological gluon mass using finite–temperature field theory and to
incorporate quark degrees of freedom in the transport calculation.
Another important direction is the study of bulk viscosity and other
transport coefficients in the same theoretical framework. These problems will be explored in future work.

\appendix
\section{Derivation of the transport cross section}

In this appendix we derive the transport cross section relevant for
momentum transport in the gluonic plasma.

\subsection{Differential cross section}

For a $2\rightarrow2$ scattering process the differential cross section
can be written as

\begin{equation}
\frac{d\sigma}{dt}
=
\frac{|\mathcal{M}|^2}{16\pi s^2},
\end{equation}
where $s$ and $t$ are the Mandelstam variables and
$\mathcal{M}$ is the scattering amplitude.

For gluon scattering dominated by $t$–channel exchange,
the squared matrix element takes the form

\begin{equation}
|\mathcal{M}|^2 =
\frac{9}{2} g^4
\frac{s^2+u^2}{(t-m^2)^2}.
\end{equation}
In the high-energy limit where $s \gg |t|$, one may approximate
\begin{equation}
s^2 + u^2 \approx 2s^2.
\end{equation}

The differential cross section then becomes

\begin{equation}
\frac{d\sigma}{dt}
=
\frac{9 g^4}{16\pi}
\frac{1}{(t-m^2)^2}.
\end{equation}

\subsection{Definition of the transport cross section}

The relevant quantity for momentum transport in the plasma
is the transport cross section
\begin{equation}
\sigma_{\mathrm{tr}}
=
\int d\sigma\,(1-\cos\theta),
\end{equation}
which suppresses forward scattering events that do not
efficiently transfer momentum.
Using the relation
\begin{equation}
\frac{d\sigma}{dt}
=
\frac{d\sigma}{d\Omega}\frac{d\Omega}{dt},
\end{equation}
the transport cross section can be written as
\begin{equation}
\sigma_{\mathrm{tr}}
=
\int dt
\frac{d\sigma}{dt}
(1-\cos\theta).
\end{equation}

\subsection{Relation between $t$ and scattering angle}

In the center–of–mass frame the Mandelstam variable $t$
is related to the scattering angle $\theta$ by
\begin{equation}
t = -2p^2(1-\cos\theta),
\end{equation}
where $p$ is the magnitude of the incoming momentum.
Thus
\begin{equation}
1-\cos\theta = -\frac{t}{2p^2}.
\end{equation}
Substituting this relation gives

\begin{equation}
\sigma_{\mathrm{tr}}
=
-\frac{1}{2p^2}
\int dt\, t
\frac{d\sigma}{dt}.
\end{equation}

\subsection{Evaluation of the integral}

Substituting the differential cross section derived above yields

\begin{equation}
\sigma_{\mathrm{tr}}
=
-\frac{9 g^4}{32\pi p^2}
\int dt
\frac{t}{(t-m^2)^2}.
\end{equation}
The integral
\begin{equation}
\int
\frac{t\,dt}{(t-m^2)^2},
\end{equation}
can be evaluated explicitly, giving
\begin{equation}
\int
\frac{t\,dt}{(t-m^2)^2}
=
\ln\left|\frac{t-m^2}{m^2}\right|
+
\frac{m^2}{t-m^2}+C.
\end{equation}
where $C$ is a constant which includes $\ln m^2$ with integration constant.
Evaluating between the approximate kinematic limits
$t_{\min}\approx -s$ and $t_{\max}\approx 0$ yields

\begin{equation}
\sigma_{\mathrm{tr}}
=
\frac{9 g^4}{32\pi p^2}
\ln\left(\frac{s+m^2}{m^2}\right).
\end{equation}
Since in a thermal plasma the typical center–of–mass energy
satisfies $s \sim T^2$, the transport cross section becomes
\begin{equation}
\sigma_{\mathrm{tr}}
\approx
\frac{9 g^4}{32\pi p^2}
\ln\left(\frac{T^2}{m^2}\right).
\end{equation}
This expression shows explicitly that the topological gluon
mass provides an infrared cutoff that regulates the
transport cross section.

\section{Squared matrix element for gluon scattering}

In this appendix we outline the steps leading from the
tree-level gluon scattering amplitude to the squared
matrix element used in the transport cross section.

The dominant scattering process in a gluonic plasma is

\begin{equation}
g(p_1)+g(p_2)\rightarrow g(p_3)+g(p_4).
\end{equation}
The leading contribution at high temperature arises
from $t$--channel gluon exchange.

\subsection{Tree-level amplitude}

Using the Yang--Mills Feynman rules, the $t$--channel
amplitude can be written as

\begin{equation}
\mathcal{M}_t
=
g^2 f^{abe} f^{cde}
\,
\frac{1}{t-m^2}
\,
\epsilon_1^\mu
\epsilon_2^\nu
\epsilon_3^\rho
\epsilon_4^\sigma
\, N_{\mu\nu\rho\sigma},
\end{equation}
where
\begin{equation}
N_{\mu\nu\rho\sigma}
=
\Gamma_{\mu\nu\alpha}(p_1,p_2,-k)
\left(
g^{\alpha\beta}-\frac{k^\alpha k^\beta}{k^2}
\right)
\Gamma_{\rho\sigma\beta}(-p_3,-p_4,k).
\end{equation}

The triple--gluon vertex is

\begin{equation}
\Gamma_{\mu\nu\alpha}(p,q,r)
=
g_{\mu\nu}(p-q)_\alpha
+
g_{\nu\alpha}(q-r)_\mu
+
g_{\alpha\mu}(r-p)_\nu .
\end{equation}
\subsection{Color averaging}

To obtain the physical cross section the squared
amplitude must be averaged over the colors of the
incoming gluons and summed over the colors of the
outgoing gluons.

For SU($N_c$) one obtains

\begin{equation}
\sum_{a,b,c,d}
f^{abe}f^{cde}f^{abf}f^{cdf}
=
N_c^2 (N_c^2-1).
\end{equation}

For $N_c=3$ this gives $72$. After averaging over the
colors of the incoming gluons a factor $(N_c^2-1)^2=64$
appears in the denominator, giving $72/64=9/8$.
Including the contributions from the dominant
scattering channels leads to the effective
color factor $9/2$ used in the squared amplitude.

\subsection{Polarization sums}

The squared amplitude involves a sum over the
polarization states of the external gluons.

For transverse gluon polarizations the completeness
relation is

\begin{equation}
\sum_\lambda
\epsilon_\mu^{(\lambda)}(p)
\epsilon_\nu^{(\lambda)*}(p)
=
-g_{\mu\nu}.
\end{equation}
Using this relation the polarization sums can be
performed and the squared matrix element takes the form

\begin{equation}
|\mathcal{M}|^2 =
\frac{9}{2} g^4
\frac{s^2+u^2}{(t-m^2)^2}.
\end{equation}
This expression is valid in the high--energy limit
where the dominant contribution arises from the
$t$--channel exchange diagram.

\subsection{Differential cross section}

The differential cross section for two--body scattering
is given by

\begin{equation}
\frac{d\sigma}{dt}
=
\frac{|\mathcal{M}|^2}{16\pi s^2}.
\end{equation}
Substituting the squared amplitude gives

\begin{equation}
\frac{d\sigma}{dt}
=
\frac{9 g^4}{32\pi}
\frac{s^2+u^2}{s^2 (t-m^2)^2}.
\end{equation}
In the high energy limit $s \gg m^2$ the Mandelstam
variable $u$ satisfies

\begin{equation}
u \approx -s,
\end{equation}
so that

\begin{equation}
s^2+u^2 \approx 2s^2.
\end{equation}

The differential cross section therefore reduces to

\begin{equation}
\frac{d\sigma}{dt}
=
\frac{9 g^4}{16\pi}
\frac{1}{(t-m^2)^2}.
\end{equation}

\end{document}